# Alloying induces directionally-dependent mobility and alters migration mechanisms of faceted grain boundaries

Megan J. McCarthy [a], Timothy J. Rupert [a,b,*]
[a] Materials and Manufacturing Technology, University of California, Irvine, CA 92697, USA
[b] Department of Materials Science and Engineering, University of California, Irvine, CA 92697, USA
* Email: trupert@uci.edu

Faceted grain boundaries exhibit unusual segregation and migration tendencies. To gain a deeper understanding of how solute atoms interact with faceted interfacial structures during migration, this study probes the migration behavior of a faceted Σ11 boundary in Cu doped with Ag atoms. The solutes are found to segregate to the facet with more free volume and strongly reduce boundary velocity in one migration direction, but not the other, due to the presence of a directionally-dependent motion mechanism that can escape solute pinning and therefore speed up migration. Hence, a new mechanism of chemically-induced anisotropy in grain boundary mobility is uncovered by these simulations.

Grain boundary segregation engineering has proven to be a powerful means of microstructural and material property manipulation [1]. Faceted grain boundaries are a promising subset of interfaces to potentially utilize in segregation engineering due to their complex, yet highly ordered structures, which classically consist of a low energy plane, a high energy plane, and facet junctions. All three structures have been shown to be preferred sites of impurity and dopant segregation [2-4]. In addition, Peter et al. [5] have shown that segregation can even induce nano-faceting in initially flat boundaries. A study of how dopants modulate the dynamic behavior of faceted boundaries is of great fundamental and technical importance, especially given recent discoveries of their unusual motion mechanisms and mobility trends [6-8]. In particular, faceted $\Sigma 11$ boundaries offer a unique opportunity to explore both segregation and migration, given their highly unusual geometry and special properties that have captured the attention of researchers for decades [9-14]. In this work, the segregation and migration behavior of a faceted $\Sigma 11$ boundary is studied with atomistic modeling techniques, using Cu as the matrix element and Ag as the solute.

All atomistic simulations were run using the LAMMPS software package [15] with visualization and post-processing conducted using the OVITO software [16]. Molecular dynamics (MD) simulations were run in the NPT ensemble with an integration timestep of 1 fs and at zero pressure. To create the faceted sample, a bicrystal-generating algorithm created by Tschopp et al. [17] was used with an embedded-atom method interatomic potential designed for the Cu-Ag system [18]. This alloy was selected because Ag has been shown to segregate the boundary [19]. Initial testing confirmed that pure Cu $\Sigma 11$ boundaries also facet with this potential, matching simulations of the boundaries with dedicated pure Cu potentials [13, 20]. After selecting a cell size that fits fully periodic boundary conditions for a given bicrystal orientation, the two grains are shifted relative to each other in small increments. The cell with the lowest grain boundary energy (averaged over the two boundaries in each bicrystal sample) is then chosen and replicated to create the simulation cell, with dimensions of 180 x 381 x 36.5 Å (*X, Y, Z* respectively) and ~200,000 atoms. To relax the boundaries, a random velocity is assigned to atoms to initialize temperature and the cell is then ramped to the target temperature over approximately 20 ps, to be annealed for a further 100 ps.

Two periodically-repeating units of the faceted $\Sigma 11$ boundary studied here, with a boundary plane inclination angle, $\beta$, at 15.8º, are shown in Fig. 1(a). Atoms are colored by their local crystal orientation using the Common Neighbor Analysis method in OVITO [16] (green atoms are face-centered cubic (FCC), red atoms are hexagonal close-packed (HCP), and white atoms have no identifiable crystal structure). The ascending plane (left-to-right in the positive X-direction) is a facet oriented along the symmetric boundary plane (SBP) of $\Sigma 11$ ($\beta = 0º$). As shown in a schematic overlayed onto one facet period in Fig. 1(b), the SBP is characterized by a chain of diamond-shaped structural units (outlined in black). The descending plane linking each symmetric facet corresponds to an incommensurate boundary plane (IBP) oriented along $\{111\}_A/\{001\}_B$. The IBP facets have a significant amount of excess free volume compared to the SBP facets, due to the presence of the $\{001\}$ plane, as can be seen when looking at the mean Voronoi volume (averaged at each X-position) in Fig. 1(c) (including FCC atoms). Emerging from each IBP facet are small structures marked by black arrows in Fig. 1(a), which are Shockley partial dislocations. This phenomenon is called grain boundary stacking fault emission, which is a common interfacial relaxation mechanism in low stacking fault energy materials such as Au and Cu, especially in boundaries with large amounts of free volume [10, 21-23]. Their dislocation character can be observed in the hydrostatic stress field in Fig. 1(d), which shows a tension-compression discontinuity (lighter colors to dark blue, respectively) characteristic of dislocation cores. Because of their distinctive nature and important role in migration, we call the sites of Shockley emission facet *nodes* and these sites can define the periodicity of the facet pattern.

A hybrid Monte Carlo (MC)/MD algorithm was used to dope the pure Cu bicrystal with Ag. Before doping, fully-minimized pure samples were annealed as described above. Configurations of Ag atoms corresponding to the target concentration were then sampled by performing 1 MC step for every 100 MD steps. MC steps were conducted using a variance-constrained semi-grand canonical ensemble [24]. The 20 ps-averaged change in the absolute value of the potential energy gradient was monitored until it reached a value less than 0.1 eV/ps, then run for 1 ns longer to generate different (but energetically equivalent) configurations of the doped bicrystals.

Fig. 2(a) contains a map of the free volume in and around the pure boundary from Fig. 1, which corresponds well to the hydrostatic stress data of Fig. 1(d). The rows of Fig. 2(b) show snapshots of the boundary with increasing Ag concentrations at 300 K, with the boundary plane indicated by dashed lines to guide the eye. At this temperature, virtually no Ag atoms are left in the bulk, consistent with the positive enthalpy of segregation for Ag in Cu [19]. As shown in the snapshot for 0.1 at.% Ag, atoms segregate first to the IBP facet, specifically to sites of highest positive free volume. These are also generally the sites of the largest positive hydrostatic stresses, as shown in Fig. 1(d). This is consistent with the fact that Ag atoms are larger than Cu atoms and therefore prefer to segregate to interfacial sites under tension. Though out of the scope of this study, one unique possibility for modulating the number and strength of segregation sites in faceted boundaries is through variation of $\beta$, which can lead to significant changes in the facet periodicity and thus overall energy [13, 20].

For concentrations above 0.1 at.% Ag, it is clear that the IBP facets are the preferred sites of segregation, with SBP facets only being occupied after the IBP facets have been saturated. This trend reflects known relationships between grain boundary energy anisotropy and segregation [25] and can also be explained by the positive enthalpy of mixing of Ag in Cu, which promotes clustering of Ag atoms. Interestingly, at 2.0 at.-% Ag there remain sites of low or no solute occupation (black arrows) at the regions on the compressive side of the partial dislocation stress field (dark blue in Fig. 2(a)). The range of behaviors observed in this boundary underscores the need for nuanced models of interfacial segregation. For example, faceted boundaries such as these present a clear example of a case where a single segregation energy for Cu-Ag or even for $\Sigma 11$ itself is not an adequate description. In the context of a polycrystal, the influence of topologically complex boundaries like these faceted interfaces could be best captured by treating segregation energy as a spectrum, rather than an average, as explored in recent works by Wagih and Schuh [26, 27].

Fig. 2(c) shows changes in segregation while varying temperature for the 0.5 at.% Ag sample, with this concentration chosen as it appears to be near the Ag-saturation limit of IBP facets, making the effect of changes in temperature clearer to observe. The rows of Fig. 2(c) present atomic snapshots with local Ag composition along the Y-direction shown to the left and along the X-direction to the right. All spatial composition plots show an average taken over 100 snapshots from different MD steps of energetically equivalent configurations. The data show that dissolution of clustered Ag atoms begins above 500 K. At and below this temperature, samples have peak concentrations of ~20 at.% in the Y-direction, at the grain boundary, and peak concentrations of ~10 at.% in the X-direction, along the IBP facets. An increase in temperature to 700 K begins dissolution of clustered Ag atoms, which reduces the peak in the Y-direction to approximately 11.5 at.% and the IBP facet peaks in the X-direction from approximately 9 at.% to 5 at.% Ag. At 900 K and higher, the Y-axis peak has dropped to under 2.5 at.% Ag and the X-axis peak is only slightly enriched (0.7-1.2 at.% Ag) above the bulk composition of ~0.4 at.% Ag. Mobility studies are run on the final configuration at 1085 K, outlined in black.

As demonstrated in Fig. 2, the complex topology of faceted boundaries determines their segregation patterns. For the same reason, their migration differs significantly from that of flat boundaries [6-8]. Each plane or defect in a faceted boundary may thus have its own unique response to the presence of dopant atoms, whether initially segregated or encountered during migration through the bulk crystal. With these two situations in mind, the specimen with 0.5 at.% Ag at 1085 K (corresponding to a homologous temperature of 0.8) was ultimately chosen for mobility studies. As can be seen in the bottom panels of Fig. 2(c), this configuration allows for slight boundary segregation (to influence structure) without excessive boundary pinning (which could render the interface immobile). Mobility studies were performed using the ECO artificial driving force (ADF) code by Ulomek et al. [28]. The growth of Grain A at the expense of Grain B (boundary motion in the negative Y-direction in all snapshots) is called *Type A* migration, and its opposite is called *Type B* migration. For each configuration, 10 unique simulations were run for at least 150 ps using an added energy value of 25 meV/atom. Doped samples used identical starting configurations, but unique velocity seeds, for Type A/B motion to understand the influence of solutes on directional migration. Because some boundaries exhibited an initial lag time before migrating (especially Type B), all measurements were taken after boundaries had moved at least 3 Å, which is then chosen to represent $t = 0$ ps. Boundary motion was tracked by first lightly minimizing the bicrystal to remove thermal noise, then locating the mean position of non-crystalline (i.e. grain boundary) atoms.

The resulting average trajectories of the pure and doped samples are shown in Figs. 3(a) and (b), respectively. The colored regions surrounding each curve shows the standard deviation accounting for the 10 different runs. Mobility, *M*, is calculated from these trajectories using the formula $M = v/P$, where *v* is velocity and *P* is the driving force. The pure samples have similar trajectories for Type A (red) and Type B (blue) migration, with average mobilities of 29.2 m·s/GPa and 32.3 m·s/GPa, respectively. In contrast, the doped Type A/B samples have distinctly different trajectories from each other. Compared to the pure Type B boundary, the doped Type B mobility is significantly slower at 12.5 m·s/GPa. The doped Type A mobility, measured at 30.5 m·s/GPa, is approximately the same as that of the pure boundaries, but 2.4 times faster than the doped Type B migration. Although the larger standard deviation in the doped Type A-migrating boundaries compared to the pure Type A indicates that they too are affected by dopants, only the doped Type B-migrating ones appear to be systematically affected.

The significant difference between Type A/B migration means that the doped bicrystals are an example of *directionally-anisotropic mobility*, or boundary mobilities that vary with the direction of motion. This behavior has been observed in faceted $\Sigma 11$ boundaries in pure Cu and Ni with different grain boundary plane orientations [20, 29]. It was found that both Type A/B-driven boundaries migrate via a series of transformations at nodes and facets. One of these transformations involves the dissociation of atomic columns at the node into a mobile clump of atoms, called *disordered clusters*. Their migration resembles the string-like collective motion reported by Zhang and coworkers [30-32]. An example of two such clusters undergoing migration is shown in Figs. 3(c) and (d). In addition, another migration mechanism called *slip plane shuffling (SPS)* can operate simultaneously during Type A motion, which involves structural transformations of the IBP facets.[29] The additional migration pathways available due to SPS can lead to larger Type A mobilities over Type B, and thus directionally-anisotropic mobility. A more detailed treatment of each mechanism and a discussion of how they compete across multiple bicrystal configurations in pure materials can be found in Refs. [20, 29].

The influence of the SPS mechanism can be illustrated by a node that is initially strongly pinned by dopant adsorption, shown in Fig. 3(e). The black arrow indicates the node location, while the dashed black line shows the orientation of the IBP plane and also acts as a fiducial marker in the following panels. As shown in Fig. 3(f), application of the Type B driving force for 100 ps does not result in any node migration. In contrast, the Type A driving force applied to the same structure (Fig. 3(g)) results in almost immediate migration. By 30 ps, the node has successfully migrated several Å to the lower left corner, which has in the process lengthened the SBP facet and created two new stacking faults in Grain A. Incidentally, analysis of this defect and very similar ones in pure boundaries with the Dislocation Analysis Algorithm in OVITO [33] reveals them to be Lomer-Cotrell locks, with two stacking faults terminating in a sessile stair rod dislocation (purple arrow). Such Lomer-Cotrell locks provide a feature which will remain in the microstructure and provide a target for experimental characterization in future work. Though this exact example comes from the starting configuration and would not alter the measured Type B mobility, only add a lag time, it is instructive for visualizing SPS and this same mechanism is observed to move the boundary past subsequent obstacles during Type A motion through the simulation cell. In contrast, nodes undergoing Type B migration have only movement enabled by disordered clusters and are fully dependent on this mechanism for moving past obstacles. Because these mechanisms operate locally, and nodes migrate generally independently of each other, this observation suggests that chemically-induced directionally-anisotropic mobility should operate similarly in many different contexts (i.e., in polycrystalline systems and with variations in solute atom type and concentration).

To understand whether disordered cluster motion is affected by alloying, an algorithm for identifying and characterizing disordered clusters during migration was applied. The distribution of grain boundary atom potential energies and the OVITO Cluster Analysis Algorithm [34] were used to identify and spatially sort mobile clusters (an example of a resulting identification was shown in Fig. 3(d)). The results of this analysis are presented in Fig. 4, with the data normalized into per node values. Starting with the average cluster size data in Fig. 4(a), disordered clusters during doped Type B motion are significantly smaller than those seen in the other three configurations, which are almost identical to each other, mirroring the mobility values from Figs. 3(a) and (b). A look at the average number of clusters per node in Fig. 4(b) further underscores the importance of disordered cluster size, rather than frequency, for migration. Interestingly, disordered cluster counts are significantly higher for doped boundaries in absolute terms, but doped Type A migration results in very similar mobilities to those of pure Type A/B-migrating boundaries. Therefore, cluster size is a more important parameter for boundary mobility than the number of clusters formed. Taken together, these plots suggest that dopant atoms act as sites of initial cluster nucleation (hence the higher overall counts in Fig. 4(b)) but may also interrupt mechanisms that would increase their size, for example, by interfering with the transport of excess free volume within the cluster [31, 32, 35].

In summary, we have explored the effects of Ag segregation on the structure and migration of a faceted $\Sigma 11$, $\beta = 15.8º$ boundary in Cu. It is shown that segregation to the facet with more free volume is preferred, with solute atoms remaining concentrated at those sites even up to homologous temperatures of 0.8. Migration studies at high temperature reveal that solute atoms strongly affect boundary velocity only in one motion direction, leading to *directionally-anisotropic mobility*. This behavior arises from the operation of the SPS facet migration mechanism only possible during Type A motion, which allows migrating facets to escape solute pinning. SPS also results in grain boundary migration-generated Lomer-Cotrell locks in both pure and doped boundaries. This study demonstrates that grain boundary segregation can lead to unexpected migration behavior.

**Acknowledgements**

This study was supported by the U.S. Department of Energy, Office of Basic Energy Sciences, Materials Science and Engineering Division under Award No. DE-SC0021224.

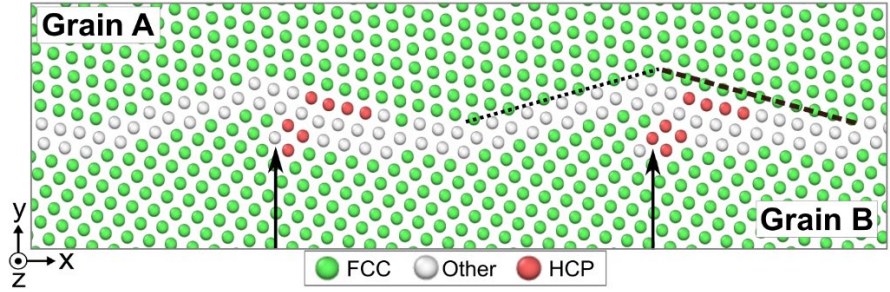

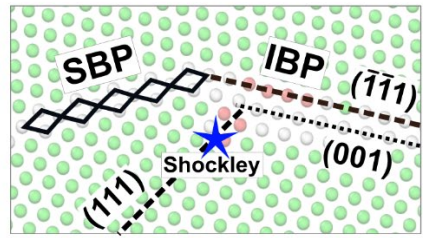
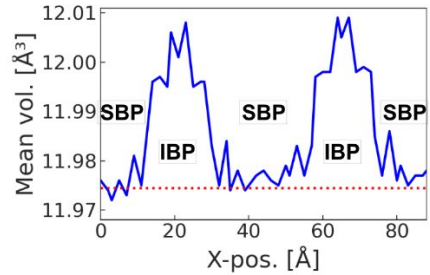

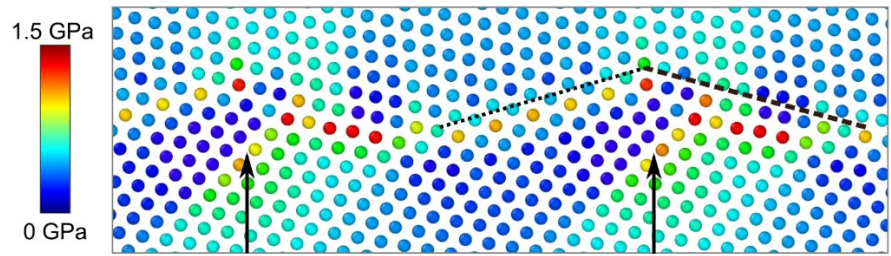

Fig. 1. (a) The faceted Σ11 boundary in pure Cu at 300 K, where the black arrows indicate the facet nodes. The ascending dotted line shows the facet oriented along the symmetric boundary plane (SBP), and the descending dashed line the faceted oriented along the $\{111\}_A/\{001\}_B$ incommensurate boundary plane (IBP). (b) Schematic overlay showing important structural features of one facet period. The nodes coincide with Shockley partials emitted from the boundary. (c) The mean Voronoi volume as a function of X-position, showing the different volume contributions of each facet. The red dotted line indicates the mean for FCC atoms alone (11.975 Å³). (d) The atomic hydrostatic stress, showing that the IBP facet has the highest tensile stresses (red), and the emitted stacking fault at the node contains the largest stress discontinuity (lacking a smooth gradient in color), characteristic of dislocation cores. The legend has been truncated to increase contrast.

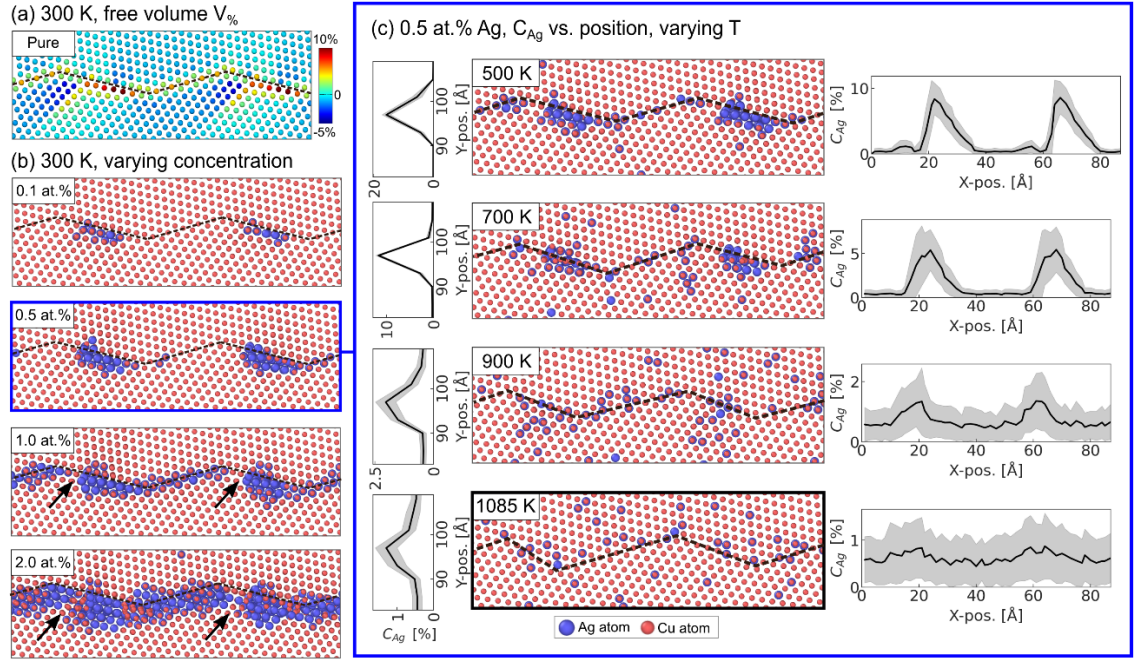

Fig. 2. (a) Faceted Σ11 boundary at 300 K, colored by free volume, where the dashed lines indicate the boundary position. (b) Atomic images at 300 K with increasing Ag concentration from 0.5 at.% to 2.0 at.% Ag. The sites of largest negative free volume remain relatively free of Ag for 1.0 and 2.0 at.% (black arrows). (c) The effect of increasing temperature for a constant concentration of 0.5 at.% Ag, with spatial composition plots for the Y-direction shown to the left of each snapshot and for the X-direction to the right for each. The final configuration chosen for the mobility studies at 1085 K is outlined in black.

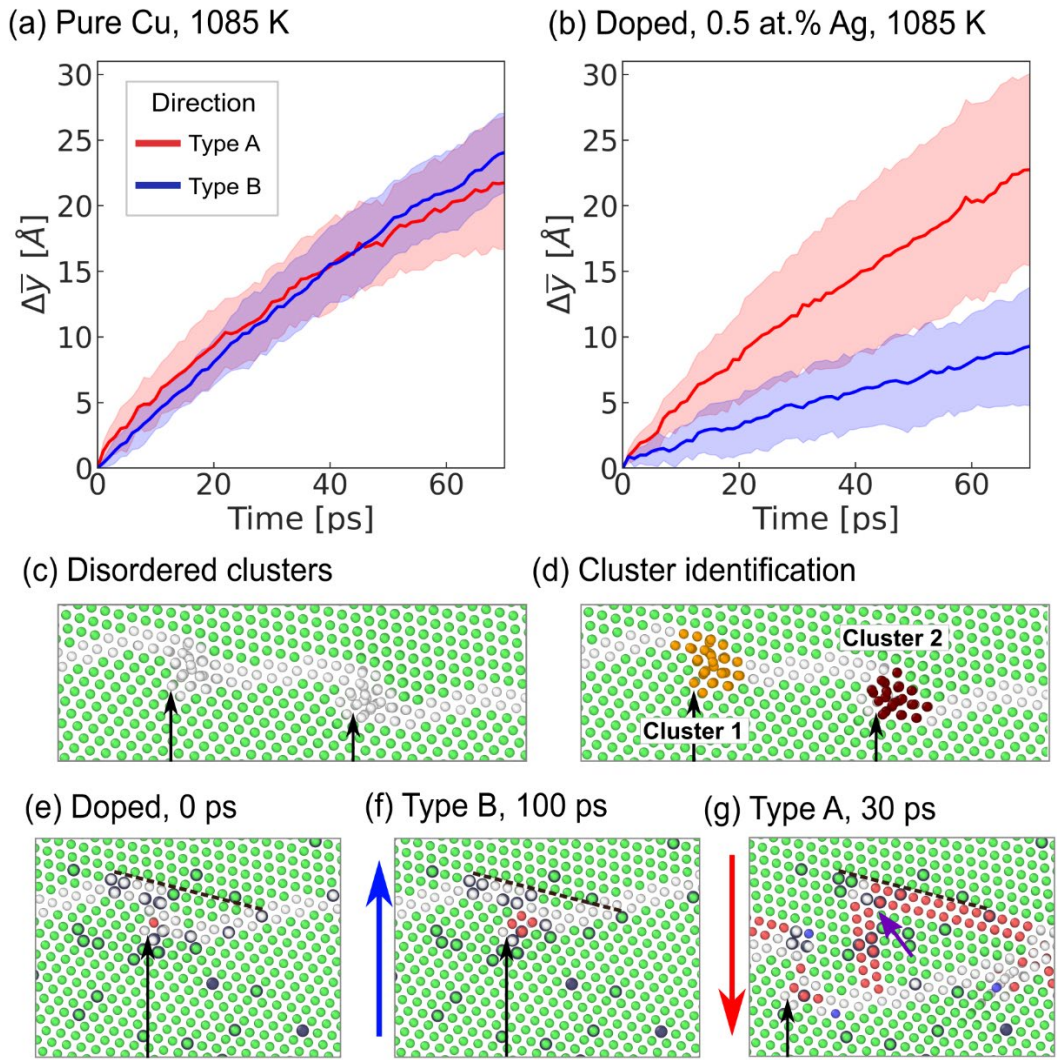

Fig. 3. (a, b) Mean boundary displacement as a function of time for Type A (red) and Type B (blue) migration for 10 runs each. The two curves of the pure boundaries in (a) overlap, while those in (b) are distinctly different from one another, indicating directionally-anisotropic mobility for the doped samples. (c, d) Examples of disordered clusters that form during node migration. (e) Example of a strongly-pinned facet node before migration, where the dashed black line in Grain A is a fiducial marker oriented along the IBP. (f) Application of the Type B ADF for 100 ps does not result in any node migration. (g) Application of the Type A ADF migrates this segment significantly in 30 ps and also forms a Lomer-Cotrell lock (purple arrow).

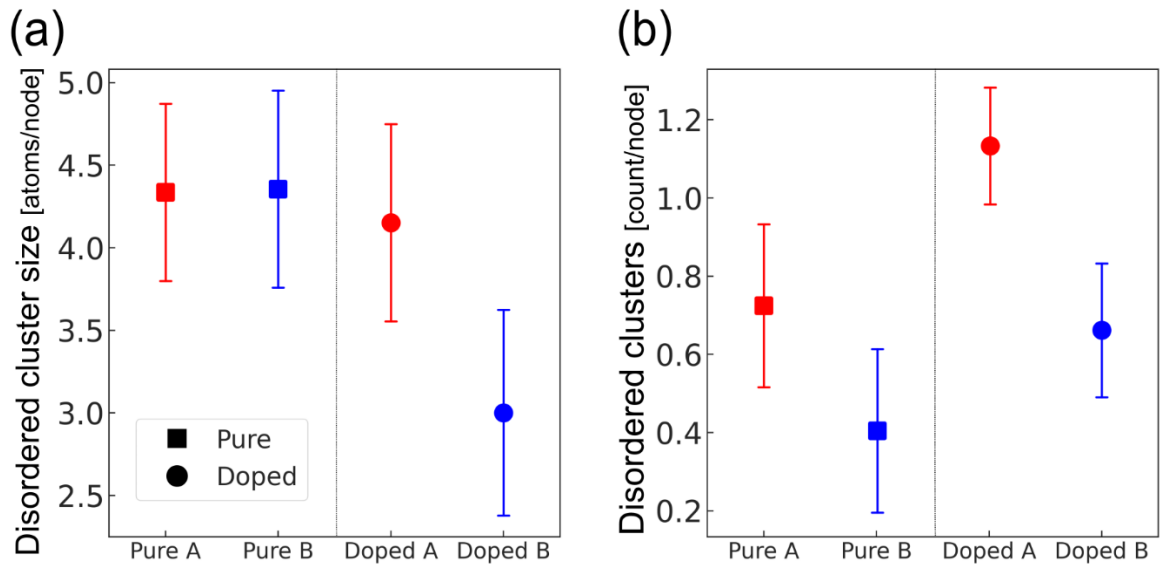

Fig. 4. (a) The average disordered cluster sizes measured during migration. The pure Type A/B and the doped Type A sizes are very similar to one another, while the clusters in the doped Type B sample are considerably smaller. (b) The average number of disordered clusters counted per node during migration, where a systematic difference between Type A/B counts in both pure and doped boundaries is observed.